\documentclass[prd,twocolumn,superscriptaddress,amsmath,amssymb,%
nofootinbib]{revtex4-1}

\begin{document}

\title{Microscopic entropy of higher-dimensional nonminimally dressed
Lifshitz black holes}

\author{Eloy Ay\'on-Beato}
\email{ayon-beato-at-fis.cinvestav.mx}
\affiliation{Departamento de F\'{\i}sica, CINVESTAV--IPN,
Apdo.\ Postal 14--740, 07000, CDMX, M\'exico}
\author{Mois\'es Bravo-Gaete}
\email{mbravo-at-ucm.cl}
\affiliation{Facultad de Ciencias B\'asicas,
Universidad Cat\'olica del Maule, Casilla 617, Talca, Chile}
\author{Francisco Correa}
\email{francisco.correa-at-uach.cl}
\affiliation{Instituto de Ciencias F\'{\i}sicas y Matem\'aticas,
Universidad Austral de Chile, Casilla 567, Valdivia, Chile}
\author{Mokhtar Hassaine}
\email{hassaine-at-inst-mat.utalca.cl}
\affiliation{Instituto de Matem\'atica y F\'isica,
Universidad de Talca, Casilla 747, Talca, Chile}
\author{Mar\'ia Montserrat Ju\'arez-Aubry}
\email{mjuarez-at-astate.edu}
\affiliation{Arkansas State University,
Carretera estatal 100, km 17.5, Municipio Col\'on, 76270, Quer\'etaro,
M\'exico}

\begin{abstract}
In arbitrary dimension, we consider a theory described by the most general
quadratic curvature corrections of Einstein gravity together with a
self-interacting nonminimally coupled scalar field. This theory is shown to
admit five different families of Lifshitz black holes dressed with a
nontrivial scalar field. The entropy of these configurations is
microscopically computed by means of a higher-dimensional anisotropic
Cardy-like formula where the role of the ground state is played by the
soliton obtained through a double analytic continuation. This involves
calculating the correct expressions for the masses of the
higher-dimensional Lifshitz black hole as well as their corresponding
soliton. The robustness of this Cardy-like formula is checked by showing
that the microscopic entropy is in perfect agreement with the gravitational
Wald entropy. Consequently, the calculated global charges are compatible
with the first law of thermodynamics as well as an anisotropic
higher-dimensional version of the Smarr formula. Some of these
configurations exist on Lifshitz critical points of the theory where all
their extensive thermodynamic quantities vanish.
\end{abstract}

\maketitle
%\tableofcontents

%%%%%%%%%%%%%%%%%%%%%%%
\section{Introduction}
%%%%%%%%%%%%%%%%%%%%%%%

Gauge/gravity duality can be extended to nonrelativistic systems by using
anisotropic spacetimes. In this context, the archetypal example is the
Lifshitz spacetime \cite{Kachru:2008yh}
\begin{equation}\label{eq:Lifshitz}
ds^2=-\frac{r^{2z}}{l^{2z}}dt^2+\frac{l^2}{r^2}dr^2
+\frac{r^2}{l^2}\sum_{i=1}^{D-2}dx_{i}^{2},
\end{equation}
whose main feature is the isometry allowing time and space to scale with
different exponents. Here, $z$ is the  dynamical critical exponent
responsible for the anisotropic scaling characterizing nonrelativistic
systems.

As it was preliminarily emphasized in \cite{Kachru:2008yh}, standard vacuum
Einstein gravity cannot allow Lifshitz spacetimes, except in the isotropic
case $z=1$ where they turn out to be anti-de~Sitter (AdS) spaces.
Nevertheless, this problem can be circumvented by considering instead
higher-order corrections to gravity theories or by introducing specific
matter sources. It then becomes important to find specific gravity models
that can accommodate the Lifshitz spacetimes together with their black hole
extensions recovering the anisotropic scaling asymptotically. These so-called
Lifshitz black holes are supposed to holographically capture the
finite-temperature behavior of their strongly correlated nonrelativistic dual
systems. New Massive Gravity \cite{Bergshoeff:2009hq} was one of the first
gravity models that was shown to admit an analytic Lifshitz black hole as
part of its vacua \cite{AyonBeato:2009nh}, which is a property that later
resulted to be generic for higher-order pure gravity theories in higher
dimensions \cite{AyonBeato:2010tm}. In presence of specific matter sources,
Lifshitz solutions have also been investigated, see
e.g.~\cite{Azeyanagi:2009pr,Giacomini:2012hg,Bravo-Gaete:2013dca,%
Correa:2014ika,Herrera:2017ztd}. Also, charged Lifshitz solutions can be
engineered through a Maxwell-Proca model \cite{Pang:2009pd}, its nonlinear
generalization \cite{Alvarez:2014pra} or in the presence of dilaton couplings
\cite{Taylor:2008tg,Tarrio:2011de}.

The relevance of Lifshitz black holes lies in the hope that strongly coupled
condensed matter systems can be better understood at finite temperature from
a holographic point of view. But because of their unconventional asymptotic
behavior, these black holes present interesting features which deserve more
profound investigations. For example, their thermodynamic properties are
usually quite different from those of the isotropic AdS black holes and in
particular if the solutions are charged, see
e.g.~\cite{Bravo-Gaete:2015xea,Zangeneh:2015uwa}. On the other hand,
three-dimensional configurations are usually excellent laboratories to
investigate important conceptual questions about the gauge/gravity duality.
For example, it has been shown that the semiclassical entropy of
three-dimensional black holes with Lifshitz asymptotics can be recovered
through a Cardy-like formula where the mass of their corresponding Lifshitz
solitons explicitly appears, giving a prominent role to these regular
configurations \cite{Gonzalez:2011nz}. The solitons are obtained from the
black holes by means of a double Wick rotation that involves inverting the
dynamical critical exponent, as result they enjoy the same sort of uniqueness
as the black holes \cite{Ayon-Beato:2014wla}. The robustness of this formula
has been successfully tested in a system exhibiting a wide spectrum of
Lifshitz configurations as is the case of self-interacting scalar fields
nonminimally coupled to New Massive Gravity \cite{Ayon-Beato:2015jga}.
Recently, this Cardy-like formula has been extended to higher-dimensional
anisotropic black holes \cite{Shaghoulian:2015kta, Shaghoulian:2015lcn,
BravoGaete:2017dso}. In the present work, we pretend to test the validity of
this higher-dimensional Cardy-like formula by considering again
self-interacting scalar fields, but nonminimally coupled now to the most
general quadratic curvature corrections of Einstein gravity in higher
dimensions. We hope this study will contribute to highlight the importance of
the role played by the soliton in the description of the thermal properties
of black holes, and particulary for those with unconventional asymptotic
behaviors.

The paper is organized as follows. In the next section we will present the
theory, field equations as well as a specific ansatz allowing particular
Lifshitz black hole solutions. Using this ansatz in Sec.~\ref{sec:PreliTQ},
the thermodynamic quantities of interest as entropy, temperature and mass of
the black holes will be preliminarily evaluated together with the mass of the
solitons. In Sec.~\ref{sec:NCLifshitzBHs}, we will explicitly present the
first two concrete classes of Lifshitz black holes fitting our ansatz. For
these solutions, we will check that their gravitational entropy, calculated
with the standard Wald formula, can be correctly reproduced by means of the
Cardy-like formula. We analyze separately in Sec.~\ref{sec:CLifshitzBHs} the
other two solutions classes within the ansatz, since they exist at Lifshitz
critical points of the theory where all the extensive thermodynamic
quantities vanish, trivially satisfying all the thermodynamic relations. In
the last section, we show the existence of a fifth class of Lifshitz black
hole that is slightly different from the initial working ansatz. For this
solution, we also test the robustness of the Cardy-like formula. In all cases
we verify the fulfillment of the first law of black hole thermodynamics and
the related anisotropic version of the Smarr formula, which point to the
correctness of the quasilocal off-shell extension of the ADT formalism that
we use to compute global charges. Finally, after some reflections on
interesting aspects raised by the new configurations that we include in the
conclusions section, we provide several appendices for reporting some of the
quite involved expressions for the (coupling) constants of the specific
solutions.

%%%%%%%%%%%%%%%%%%%%%%%%%%%%%%%%%%%
\section{Setup of the problem}
%%%%%%%%%%%%%%%%%%%%%%%%%%%%%%%%%%%

In arbitrary dimension $D$, we consider a gravity action given by the most
general quadratic curvature corrections of the Einstein-Hilbert action
sourced by a self-interacting nonminimally coupled scalar field {\small
\begin{equation}\label{eq:Squad}
S[g_{\mu\nu},\Phi]=\int{d}^Dx\sqrt{-g}(\mathcal{L}_\text{g}
+\mathcal{L}_\text{s}),
\end{equation}}%
with {\small
\begin{align*}
\mathcal{L}_\text{g}&=\frac{1}{2\kappa}\left(R-2\lambda+\beta_1{R}^2
+\beta_2{R}_{\alpha\beta}{R}^{\alpha\beta}
+\beta_3{R}_{\alpha\beta\mu\nu}{R}^{\alpha\beta\mu\nu}
\right),\nonumber\\
\mathcal{L}_\text{s}&=-\frac{1}{2}\nabla_{\mu}\Phi\nabla^{\mu}\Phi
-\frac{1}{2}\xi R\Phi^2-U(\Phi).\nonumber
\end{align*}}%
The values of the cosmological constant $\lambda$, the coupling constants
$\beta_n$, the self-interacting potential $U(\Phi)$ and eventually the
nonminimal coupling parameter $\xi$ will explicitly depend on the concrete
solutions presented in the sections that follow, and in many cases they will
be given in the appendices. The field equations obtained by varying the
action with respect to the metric and the scalar field read
\begin{subequations}{\small
\begin{align}
G_{\mu\nu}+\lambda g_{\mu\nu}+K_{\mu\nu}&=\kappa T_{\mu\nu},
\label{eqmotion1}\\
\Box\Phi - \xi R\Phi &= \frac{dU(\Phi)}{d\Phi},\label{eqmotion2}
\end{align}}%
where we have defined {\small
\begin{align}
K_{\mu\nu}={}&(\beta_2+4\beta_3)\square{R}_{\mu\nu}
+\frac12(4\beta_1+\beta_2)g_{\mu\nu}\square{R}
\nonumber\\
&-(2\beta_1+\beta_2+2\beta_3)\nabla_\mu\nabla_\nu{R}
+2\beta_3R_{\mu\gamma\alpha\beta}R_{\nu}^{~\gamma\alpha\beta}
\nonumber\\
&+2(\beta_2+2\beta_3)R_{\mu\alpha\nu\beta}R^{\alpha\beta}
-4\beta_3R_{\mu\alpha}R_{\nu}^{~\alpha}+2\beta_1RR_{\mu\nu}
\nonumber\\
&-\frac12\!\left(\beta_1{R}^2+\beta_2{R}_{\alpha\beta}{R}^{\alpha\beta}
+\beta_3{R}_{\alpha\beta\gamma\delta}{R}^{\alpha\beta\gamma\delta}
\right)\!g_{\mu\nu},
\end{align}}%
and the energy-momentum tensor is given by {\small
\begin{align}
T_{\mu\nu}={}&\nabla_{\mu}\Phi\nabla_{\nu}\Phi
-g_{\mu\nu}\left(\frac{1}{2}\nabla_{\sigma}\Phi\nabla^{\sigma}\Phi
+U(\Phi)\right)
\nonumber\\
&+\xi\left( g_{\mu\nu}\Box - \nabla_{\mu}\nabla_{\nu}+G_{\mu\nu}
\right)\Phi^2.
\end{align}}%
\end{subequations}

In order to look for Lifshitz black holes, we will opt for the following
ansatz
\begin{align}
ds^2 &= - \frac{r^{2z}}{l^{2z}} f(r) dt^2 +
\frac{l^2}{r^2} \frac{dr^2}{f(r)} +
\frac{r^2}{l^2}\sum_{i=1}^{D-2}dx_{i}^{2},\nonumber\\
\Phi&=\Phi(r),\label{lifbh}
\end{align}
where the structural metric function satisfies the boundary condition
$\lim_{r\rightarrow\infty} f(r)=1$, ensuring the metric to reproduce the
Lifshitz asymptotics (\ref{eq:Lifshitz}). As shown below, four of the five
classes of Lifshitz solutions that will be presented can be generically
parameterized as
\begin{equation}\label{gensolLif}
f(r)=1-\left(\frac{r_h}{r}\right)^{\chi},\qquad
\Phi(r)=\Phi_0\left(\frac{r_h}{r}\right)^{\frac{\chi}{2}},
\end{equation}
where $\chi$ is a non-negative decay exponent modulating the Lifshitz
asymptotics, $r_h$ denotes the location of the horizon and $\Phi_0$
characterizes the positive strength of the field. The positivity of the
scalar field can be explained from the fact that for these four classes of
solutions, the discrete transformation $\Phi\mapsto-\Phi$ will be a symmetry
of the problem. The ansatz (\ref{lifbh}-\ref{gensolLif}) is also motivated by
the fact that for $\Phi_0=0$ most of the vacuum Lifshitz black hole solutions
known for the theory have precisely this form \cite{AyonBeato:2010tm}, which
also occurs for their charged extensions \cite{Bravo-Gaete:2015xea}. The
remaining solutions belong to a different class where the structural metric
function involves two different radial powers and will be presented in
Sec.~\ref{sec:lastc}.

One of the main aims of this work is to confirm the importance of the role
played by the gravitational soliton for the thermal properties of the
Lifshitz black holes. In order to achieve this task correctly, we will need
the Lifshitz soliton counterparts of the black holes
(\ref{lifbh}-\ref{gensolLif}). The solitons will be generically described  by
the following metric
\begin{equation}
d\bar{s}^2 = - \frac{\bar{r}^2}{l^2} d\bar{t}^2 +
\frac{l^2}{\bar{r}^2} \, \frac{d\bar{r}^2}{f(\bar{r})}
 +\frac{\bar{r}^{2 z}}{l^{2 z}} f(\bar{r}) d{\bar{x}}_{1}^{2}
 +\frac{\bar{r}^2}{l^2}\sum_{i=2}^{D-2} d\bar{x}_{i}^{2},
 \label{lifsoleuclidean}
\end{equation}
with
\begin{equation}
f(\bar{r})=1-\left[\left(\frac{2}{\chi}\right)^{1/z}\frac{l}{\bar{r}}
\right]^{\chi},\,
\Phi(\bar{r})=\Phi_0
\left[\left(\frac{2}{\chi}\right)^{1/z}\frac{l}{\bar{r}}\right]
^{\frac{\chi}{2}}\label{gensolLifsoliton}.
\end{equation}
The solitons are obtained from the black holes by means of a double Wick
rotation $\bar{t}=-ix_1$ and $\bar{x}_1=-it$ supplemented by an adjustment of
the horizon location
$$
r_h=l\left(\frac{2}{\chi}\right)^{\frac{1}{z}},
$$
which ensures the correct identification of its Euclidean version.

The configurations that will be described below are fully determined in terms
of three parameters, namely, the decay exponent $\chi$, the strength $\Phi_0$
and the dynamical exponent $z$. In order to simplify the discussion, we start
by evaluating first the formulas of interest as the entropy, temperature and
mass of the black hole configurations as well as the mass of the solitons for
generic values of these constants, i.e.\ not for those that actually satisfy
all the system constraints. The precise thermodynamic quantities are given
later for each genuine solution with the help of these formulas.

%%%%%%%%%%%%%%%%%%%%%%%%%%%%%%%%%%%%%%%%%%%%%%%%%%%%%%%%%%%%%%%%%%%
\section{Preliminary thermodynamic quantities \label{sec:PreliTQ}}
%%%%%%%%%%%%%%%%%%%%%%%%%%%%%%%%%%%%%%%%%%%%%%%%%%%%%%%%%%%%%%%%%%%

First of all, the Wald entropy formula \cite{Iyer:1994ys} for action
\eqref{eq:Squad} evaluated in the black hole ansatz
(\ref{lifbh}-\ref{gensolLif}) generically reads
\begin{subequations}\label{waldgensol}
\begin{equation}
\mathcal{S}_\text{W}=\frac{2\pi\Upsilon}{\kappa}
\left(\frac{r_h}{l}\right)^{D-2}\Omega_{D-2},
\end{equation}
where $\Omega_{D-2}$ represents the finite volume of the $(D-2)$-dimensional
planar base manifold and the coefficient
\begin{align}
\Upsilon\equiv{}&-\left.\kappa\,P^{abcd} \, \varepsilon_{ab} \, \varepsilon_{cd}
\right\rvert_{r = r_h}\nonumber\\
={}&1-\kappa\xi\Phi_0^2+\frac{\chi}{l^2}\bigl[
2(\chi-3z-2D+4)\beta_1\nonumber\\
&+(\chi-3z-D+2)\beta_2 + 2(\chi-3z)\beta_3\bigr],\label{eq:Upsilon}
\end{align}
\end{subequations}
with
$P^{abcd}\equiv\partial(\mathcal{L}_\text{g}+\mathcal{L}_\text{s})/\partial
R_{abcd}$ measures how the theory departs from the behavior of standard
gravity, whose areal interpretation of black holes entropy forces
$\Upsilon=1$. On the other hand, their temperature is given by
\begin{equation}
T=\frac{1}{4\pi} \frac{r_h^{z+1}}{l^{z+1}}
f^{\prime}(r_h)=\frac{\chi}{4 \pi l} \left(\frac{r_h}{l}\right)^{z}.
\label{T}
\end{equation}

In order to compute the masses of the black hole and soliton configurations
defined in Eqs.~(\ref{lifbh}-\ref{gensolLifsoliton}), we will opt for the
quasilocal formalism as defined in \cite{Kim:2013zha,Gim:2014nba}. Notice
that this formalism has proved to be well suited for correctly computing the
masses of black holes of higher-order gravity theories with rather
unconventional asymptotics, see e.g.~\cite{Ayon-Beato:2015jga}. The
quasilocal formalism is based on an off-shell prescription \cite{Kim:2013zha}
for the ADT potential \cite{ADT} which allows the following concise
expression for the conserved charge associated to a Killing vector field $k$
\begin{equation}\label{eq:charge}
Q(k)=\!\int_{\cal B}\!d^{D-2}x_{\mu\nu}\!\left(\!\Delta
N^{\mu\nu}(k)-2k^{[\mu}\!\!\int^1_0\!\!ds~\Theta^{\nu]}(k|s)\!\right)\!,
\end{equation}
where $s$ is a parameter interpolating between the solution of interest at
$s=1$ and the asymptotic one at $s=0$, the difference between their off-shell
Noether potentials is denoted by $\Delta N^{\mu\nu}(k)\equiv
N^{\mu\nu}_{s=1}(k)-N^{\mu\nu}_{s=0}(k)$ and $\Theta^{\nu}$ is the surface
term arising after varying the action. In the present case, these tensors are
given by
\begin{align*}
\Theta^\mu={}&2\sqrt{-g}\biggl(P^{\mu(\alpha
\beta)\gamma}\nabla_\gamma\delta g_{\alpha\beta} -\delta
g_{\alpha\beta}\nabla_\gamma P^{\mu(\alpha\beta)\gamma}\\
&\qquad\quad+\frac{1}{2}\,\frac{\partial \mathcal{L}_\text{s}}{\partial
\left(\partial_{\mu}\Phi\right)}\delta\Phi\biggr),\\
N^{\mu\nu}={}&2\sqrt{-g}\left(P^{\mu\nu\rho\sigma}\nabla_\rho k_\sigma
-2 k_\sigma\nabla_\rho P^{\mu\nu\rho\sigma}\right).
\end{align*}
For a timelike Killing vector field, $\partial_t=k^{\mu}\partial_{\mu}$, the
evaluation of the mass formula for action \eqref{eq:Squad} in the black hole
ansatz (\ref{lifbh}-\ref{gensolLif}) gives rise to the expression
\begin{align}
\mathcal{M}_\text{bh}(k)={}&
\left\{-4\,\Psi_{1}
+\kappa\,\Phi_0^2\left[2\left(2\,\chi+2\,z-D+2\right)\xi-\chi\right]
\right\}\nonumber\\&\times
\left(\frac{r_h}{l}\right)^{2\chi}
\left(\frac{l}{r}\right)^{2\chi-z-D+2}\frac{\Omega_{D-2}}{8\,\kappa\,l}
\nonumber\\
&+\left\{2\,\Psi_{2}
-\kappa\,\Phi_0^2\left[4(\chi+z)\xi-\chi\right]\right\}\nonumber\\&\times
\left(\frac{r_h}{l}\right)^{\chi}
\left(\frac{l}{r}\right)^{\chi-z-D+2}\frac{\Omega_{D-2}}{4\,\kappa\,l},
\label{massgensol}
\end{align}
where $\Psi_1$ and $\Psi_2$ are two dimensionless linear combinations of the
squared corrections coupling constants reported in App.~\ref{a1}. For the
soliton ansatz (\ref{lifbh}-\ref{gensolLifsoliton}) with timelike Killing
vector field $\partial_{\bar{t}}=k^{\mu}\partial_{\mu}$ the mass formula
reads
\begin{align}
\mathcal{M}_\text{sol}(k)={}&
\left\{-4\,\Xi_{1}
+\kappa\,\Phi_0^2\left[2\left(4\chi-2\,z-D+6 \right)\xi-\chi\right]
\right\}\nonumber\\&\times
\left(\frac{2}{\chi}\right)^{2\chi/z}
\left(\frac{l}{\bar{r}}\right)^{2\chi-z-D+2}
\frac{\Omega_{D-2}}{8\,\kappa\,l} \nonumber\\
&+\left\{ 2\,\Xi_{2}-\kappa\,\Phi_0^2 \left[ 4(\chi+1)\xi-\chi\right]\right\}
\nonumber\\&\times
\left(\frac{2}{\chi}\right)^{\chi/z}
\left(\frac{l}{\bar{r}}\right)^{\chi-z-D+2}
\frac{\Omega_{D-2}}{4\,\kappa\,l},\label{massgensoliton}
\end{align}
where the dimensionless coupling constants combinations $\Xi_{1}$ and
$\Xi_{2}$ are also defined in App.~\ref{a1}. For actual solutions, the mass
expressions (\ref{massgensol}) and (\ref{massgensoliton}) must be global
charges and cannot depend on the radial coordinates $r$ and $\bar{r}$,
respectively. Interestingly, this imposes constraints on the constants $z$,
$\chi$ and $\Phi_0$ giving indications on the possible solutions within the
ansatz; concretely, only two families of decay exponents $\chi$ are possible
since they are the only ones giving rise to nontrivial global charge masses.

In what follows, we will report four different classes of Lifshitz black hole
solutions fitting our ansatz (\ref{lifbh}-\ref{gensolLif}). For each
solution, we will check that its gravitational Wald entropy
(\ref{waldgensol}) is correctly reproduced by means of a higher-dimensional
anisotropic Cardy-like formula \cite{BravoGaete:2017dso} given by {\small
\begin{equation}\label{generCardyd}
\mathcal{S}_\text{C}=\frac{2\pi l(z+D-2)}{D-2}
\!\left(\!-\frac{(D-2)\mathcal{M}_\text{sol}}{z}\!\right)^{\frac{z}{z+D-2}}
\!{\mathcal{M}_\text{bh}}^{\frac{D-2}{z+D-2}}.
\end{equation}}%
This expression is the higher-dimensional extension of the one obtained for
two-dimensional Lifshitz field theory \cite{Gonzalez:2011nz}. Here we have
used the notation $\mathcal{S}_\text{C}$ for the microscopic entropy in order
to reflect that the anisotropic Cardy-like expression is \emph{a priori}
different from the gravitational Wald entropy (\ref{waldgensol}).
Nevertheless, as shown below, both entropies will coincide for the different
classes of solutions reported. For completeness, we will also verify  that
the first law of black hole thermodynamics
\begin{equation}\label{firstlaw}
d\mathcal{M}_\text{bh}=T d\mathcal{S}_\text{W},
\end{equation}
consistently holds for each Lifshitz black hole solution.

%%%%%%%%%%%%%%%%%%%%%%%%%%%%%%%%%%%%%%%%%%%%%%%%%%%%%%%%%%%%%%%%%%%%%%
\section{Noncritical Lifshitz black holes \label{sec:NCLifshitzBHs}}
%%%%%%%%%%%%%%%%%%%%%%%%%%%%%%%%%%%%%%%%%%%%%%%%%%%%%%%%%%%%%%%%%%%%%%

Here, we will present the first two classes of solutions that fit within
ansatz (\ref{lifbh}-\ref{gensolLif}) and compute their definitive
thermodynamic quantities through the preliminary formulas derived in the
previous section. All of them will turn out to be nontrivial, unlike the
critical cases that we leave for the next section, which explains why we call
the following solutions noncritical. For each solution, we will corroborate
that their Wald entropy can be reproduced from the anisotropic Cardy-like
formula (\ref{generCardyd}). We also show that the fulfillment of the first
law (\ref{firstlaw}) for all is linked with a higher-dimensional anisotropic
version of the Smarr formula.

%%%%%%%%%%%%%%%%%%%%%%%%%%%%%%%%%%%%%%%%%%%%%%%%%%%%%%%%%%%%%%%%%%%%%%%%%%%%%%
\subsection{Class with arbitrary dynamical exponent and arbitrary
nonminimal coupling parameter\label{1stclass}}
%%%%%%%%%%%%%%%%%%%%%%%%%%%%%%%%%%%%%%%%%%%%%%%%%%%%%%%%%%%%%%%%%%%%%%%%%%%%%%

The first family of solutions is obtained for the standard potential where a
mass term is supplemented by a quartic interaction and exists for arbitrary
values of the dynamical exponent $z$ and of the nonminimal coupling parameter
$\xi$. Because of cumbersome formulas, the concrete form of the potential and
the parameterizations obeying the different coupling constants as well as the
cosmological constant are reported in App.~\ref{1st}. Its line element and
the nontrivial scalar field are given by
\begin{align}
ds^2={}&-\left(\frac{r}{l}\right)^{2z}\left[1
-\left(\frac{r_h}{r}\right)^{(z+D-2)/2}\right]dt^2\nonumber\\
&+\frac{l^2}{r^2}
\left[1-\left(\frac{r_h}{r}\right)^{(z+D-2)/2}\right]^{-1}dr^2
+\frac{r^2}{l^2}\sum_{i=1}^{D-2} dx_{i}^{2},\nonumber\\
\Phi(r)={}&\biggl(\frac1{\kappa\,l^2P_5(z;\xi)}
\Bigl\{\left[3z^2+(D+2)(D-2)\right]P_{3}(z)\,l^2\nonumber\\
&-2(D-3)(D-4)(z+D-2)P_{4}(z)\beta_3\Bigr\}\biggr)^{1/2}\nonumber\\
&\times\left(\frac{r_h}{r}\right)^{(z+D-2)/4},
\label{solutionfamily1}
\end{align}
where the polynomials $P_n$  are also defined in App.~\ref{1st} together with
the remaining details of the solution.

This solution is obtained from the proposed ansatz (\ref{gensolLif}) by using
one of the only two decay exponents allowing a well-defined Lifshitz mass,
namely, $\chi=(z+D-2)/2$. The result is the higher-dimensional lifting from
$D=3$ of the black hole family with Lifshitz decay $(z+1)/2$ originally
derived in Ref.~\cite{Correa:2014ika} for  New Massive Gravity, whose
thermodynamics was studied in Ref.~\cite{Ayon-Beato:2015jga}. It is
interesting to notice that the above higher-dimensional line element has been
previously obtained also as a vacuum solution in \cite{AyonBeato:2010tm}, but
for a more restrictive election of the coupling constants. The vacuum limit
of \cite{AyonBeato:2010tm} is easily recovered by fixing the coupling
constant $\beta_3$ in solution (\ref{solutionfamily1}) in order to obtain a
vanishing scalar strength. In this sense the present solution is also a
generalization of the one of \cite{AyonBeato:2010tm} allowing the same black
hole to be dressed by a self-interacting nonminimally coupled scalar field.
Notice that the existence of nonminimal coupling is not imperative since the
solution still exists in the limit $\xi=0$.

The thermodynamic properties of the lifted configuration follow from the
following expressions, first, the preliminary formula (\ref{waldgensol})
gives the Wald entropy
\begin{align}
\mathcal{S}_\text{W}={}& \frac{2\pi\Upsilon_1}{\kappa}
\left(\frac{r_h}{l}\right)^{D-2}\Omega_{D-2},
\end{align}
where the dimensionless coefficient $\Upsilon_1$, depending on the free
coupling constants, is mutual to all the extensive thermodynamic quantities
associated with the solution and is defined in App.~\ref{App:Ecoeffs}. The
Hawking temperature (\ref{T}) in this case reads
\begin{equation}
T=\frac{z+D-2}{8 \pi l} \left(\frac{r_h}{l}\right)^{z}.
\end{equation}
The formulas (\ref{massgensol}) and (\ref{massgensoliton}) lead to the
Lifshitz black hole and soliton masses, respectively,
\begin{align}
\mathcal{M}_\text{bh}={}& \frac{(D-2)\Upsilon_1}{4\kappa}
\left(\frac{r_h}{l}\right)^{z+D-2}\frac{\Omega_{D-2}}{l},\\
\mathcal{M}_\text{sol}={}&-\frac{z\Upsilon_1}{4\kappa}
\left(\frac{4}{z+D-2}\right)^{\frac{z+D-2}{z}}\frac{\Omega_{D-2}}{l}.
\end{align}
Now, it is straightforward to verify that the Cardy-like formula
(\ref{generCardyd}) for the entropy correctly reproduces the gravitational
Wald entropy, that is $\mathcal{S}_\text{W}=\mathcal{S}_\text{C}$.

Another interesting feature of these thermodynamic quantities is that they
obey the following anisotropic higher-dimensional version of the Smarr
formula \cite{Smarr:1972kt}
\begin{equation}\label{smarrf}
\mathcal{M}_\text{bh}=\frac{D-2}{z+D-2} T\, \mathcal{S}_\text{W},
\end{equation}
which in fact is not unexpected since it is indispensable to they
consequently respect the first law (\ref{firstlaw}).

For the other decay exponent compatible with a well-defined mass,
$\chi=z+D-2$, it happens that one obtains a vanishing mass, as will be
exhibited in Sec.~\ref{sec:CLifshitzBHs}. However, there is an exception for
the critical exponent $z=D$, this is the solution we present below.

%%%%%%%%%%%%%%%%%%%%%%%%%%%%%%%%%%%%%%%%%%%%%%%%%%%%%%%%%%%%%%%%%%%%%%%%%%%%
\subsection{Solution with dynamical exponent $z=D$ \label{2stclass}}
%%%%%%%%%%%%%%%%%%%%%%%%%%%%%%%%%%%%%%%%%%%%%%%%%%%%%%%%%%%%%%%%%%%%%%%%%%%%

The second family of Lifshitz black hole solutions exists for a dynamical
exponent $z=D$ and for a nonminimal coupling parameter $\xi<(D-1)/(5D-2)$,
\begin{align}
ds^2={}&-\frac{r^{2D}}{l^{2D}}
\left[1-\left(\frac{r_h}{r}\right)^{2(D-1)}\right]dt^2\nonumber\\&
+\frac{l^2}{r^2}
\left[1-\left(\frac{r_h}{r}\right)^{2(D-1)}\right]^{-1}dr^2%\nonumber\\&
+\frac{r^2}{l^2}\sum_{i=1}^{D-2}dx_{i}^{2},\nonumber\\
\Phi(r)={}&\frac1{\sqrt{\kappa}}\sqrt{\frac{(D-2)}{D-1-(5D-2)\xi}}
\,\left(\frac{r_h}{r}\right)^{D-1}.\label{eq:familyz32D}
\end{align}
In this case, the self-interacting potential is also given by a mass term
plus a $\Phi^4$ interaction and the coupling constants $\beta_1$ and
$\beta_3$ are arbitrary
\begin{align*}
U(\Phi)={}&-\frac{(D-1)[D-1-(5D-2)\xi]}{4(D-2)l^2}\\
&\times \left\{2(D-2)\Phi^2+[D-1-(D-2)\xi]\kappa\Phi^4\right\},
\frac{}{}\\
%       & &\nonumber\\
\lambda={}&\frac{(D-1)[8D(D-3)(D-4)\beta_3-(5D-2)l^2]}{4l^4},\\
%       & &\nonumber\\
\beta_2={}&-\frac{2(5D-2)(D-1)\beta_1+4(2D+1)\beta_3-l^2}{2(D+2)(D-1)}.
\end{align*}
We remark that the restriction on the nonminimal coupling parameter
consistently includes the minimal case $\xi=0$. This solution consistently
fits the working ansatz (\ref{gensolLif}) within the other admissible family
of decay exponents $\chi=z+D-2=2(D-1)$ when the critical exponent takes the
value $z=D$. Its Hawking temperature becomes
$$
T={\frac { \left( D-1 \right)}{2 \pi l}}
\left(\frac{r_h}{l}\right)^{D},
$$
while the entropy together with the masses of the black hole and its soliton
counterpart are given by
\begin{align*}
\mathcal{S}_\text{W}&= \frac{2\pi\Upsilon_2}{\kappa}
\left(\frac{r_h}{l}\right)^{D-2}\Omega_{D-2},\\
\mathcal{M}_\text{bh}&=
\frac{(D-2)\Upsilon_2}{2\kappa}
\left(\frac{r_h}{l}\right)^{2(D-1)}\frac{\Omega_{D-2}}{l},\\
\mathcal{M}_\text{sol}&=-\frac{D \Upsilon_2}{2\kappa}
\left(\frac{1}{D-1}\right)^{\frac{2(D-1)}{D}}\frac{\Omega_{D-2}}{l},
\end{align*}
where the mutual extensive coefficient is again defined in
App.~\ref{App:Ecoeffs}. As before, one can check the validity of the first
law and the Cardy-like formula (\ref{generCardyd}), as well as of the Smarr
formula (\ref{smarrf}).

%%%%%%%%%%%%%%%%%%%%%%%%%%%%%%%%%%%%%%%%%%%%%%%%%%%%%%%%%%%%%%%%%%%%
\section{Critical Lifshitz black holes \label{sec:CLifshitzBHs}}
%%%%%%%%%%%%%%%%%%%%%%%%%%%%%%%%%%%%%%%%%%%%%%%%%%%%%%%%%%%%%%%%%%%%

In the previous section it was shown that the extensive thermodynamic
quantities share the same coefficient $\Upsilon$, whose general definition is
given in (\ref{eq:Upsilon}). Consequently, the points in the parameter space
of the theory where $\Upsilon=0$ will give rise to Lifshitz black holes with
vanishing extensive thermodynamic quantities. Notice that in the absence of a
scalar field and for isotropic scaling $z=1$ (i.e.\ for AdS black holes;
whose standard decay is $\chi=z+D-2=D-1$ to be compatible with the AdS mass)
this is precisely the case in $D=4$ of the so-called critical gravity point
\cite{Lu:2011zk,Deser:2011xc}, where $\beta_2=-3\beta_1=-l^2/2$ giving
\begin{equation}\label{eq:CriticalG}
\Upsilon=1-\frac6{l^2}(4\beta_1+\beta_2)=0.
\end{equation}
As has been highlighted, for example in \cite{Anastasiou:2017rjf}, the common
criteria that allows to generically define critical gravity points for any
theory is that the entropy as well as the global charges of their black holes
vanish on this specific region of the parameter space. In this sense,
$\Upsilon=0$ defines Lifshitz critical points of the examined theory. In this
section we present two other families fitting our ansatz (\ref{gensolLif})
which are examples of critical Lifshitz black holes, since they exist at
points where $\Upsilon=0$, sharing the peculiarity of having vanishing Wald
entropy and masses. They trivially satisfy the first law of thermodynamics as
well as the Cardy-like formula since
$\mathcal{S}_\text{W}=0=\mathcal{S}_{C}$.

The first critical solution is obtained by choosing the other decay exponent
compatible with a well-defined Lifshitz mass $\chi=z+D-2$, which gives
\begin{align}
ds^2={}&-\frac{r^{2z}}{l^{2z}}
        \left[1-\left(\frac{r_h}{r}\right)^{z+D-2}\right]dt^2\nonumber\\
&       +\frac{l^2}{r^2}
        \left[1-\left(\frac{r_h}{r}\right)^{z+D-2}\right]^{-1}dr^2
        +\frac{r^2}{l^2}\sum_{i=1}^{D-2}d{x}_i^2,\nonumber\\
\Phi(r)={}&2\sqrt{\frac{(D-2)(z-1)}{\kappa{P}_2(z;\xi)}}
          \,\left(\frac{r_h}{r}\right)^{\frac{z+D-2}{2}},
\end{align}
where the second grade polynomial in the critical exponent at the denominator
of the scalar strength is defined as
\begin{equation*}
P_2(z;\xi)\equiv(z+D-2)^2-4[2z^2+(D-2)(2z+D-1)]\xi.
\end{equation*}
This solution is the higher-dimensional lifting of the family with Lifshitz
decay $z+1$ obtained in Ref.~\cite{Correa:2014ika}. The self-interaction
potential supporting the solution and the values of the coupling constants
are extended as
\begin{align*}
U(\Phi)={}&-\frac{P_2(z;\xi)}{16l^2}\!
\left[2\Phi^2+\left(\!\frac{(z+D-2)^2}{4(D-2)(z-1)}-\xi\!\right)\!
\kappa\Phi^4\right]\!,\\
\beta_1={}&\frac{(D-2)z^2+2(D-3)z-(D-2)^2}{(D-2)(z-1)(z+D-2)}\beta_3\\
&+\frac{(1-4\xi)l^2}{2P_2(z;\xi)},\\
-\beta_2={}&\frac{(D-2)(4z^2-D^2+3D-4)+2D(D-3)z}{(D-2)(z-1)(z+D-2)}\beta_3\\
&+\frac{l^2}{2P_2(z;\xi)},
\end{align*}
while the cosmological constant $\lambda$ takes the same expression given in
App.~\ref{1st}. It is straightforward to check by means of formulas
(\ref{massgensol}-\ref{massgensoliton}) that the masses of the black hole and
its soliton counterpart are zero, which is a property also shared by the Wald
entropy (\ref{waldgensol}).

The other critical solution is obtained for the decay exponent $\chi=2(z-1)$
which does not give, in general, a global charge. However, together with its
accompanying coupling constants they exactly cancel the coefficients in front
of the decaying powers that prevent the mass formula from becoming a
conserved charge, causing its vanishing at the same time. The resulting
solution is
\begin{align}
ds^2={}&-\left(\frac{r}{l}\right)^{2z}
        \left[1-\left(\frac{r_h}{r}\right)^{2(z-1)}\right]dt^2\nonumber\\
&       +\frac{l^2}{r^2}
        \left[1-\left(\frac{r_h}{r}\right)^{2(z-1)}\right]^{-1}dr^2
       +\frac{r^2}{l^2}\sum_{i=1}^{D-2}d{x_i}^2,\nonumber\\
\Phi(r)={}&\biggl( \frac1{\kappa l^2\tilde{P}_{2}(z;\xi)}
           \bigl[4(D-3)(D-4)z(z-D)\beta_3\nonumber\\
&          -(D-1)(2z-D-2)l^2\bigr]\biggr)^{1/2}
           \left(\frac{r_h}{r}\right)^{z-1}, \label{3nd}
\end{align}
where the critical exponent polynomial is given by
\begin{equation*}
\tilde{P}_2(z;\xi)\equiv[2z^2+(D-2)(2z+D-1)]\xi-(D-1)(z-1),
\end{equation*}
and the specific parameterizations of the coupling constants together with
the cosmological one are presented in App.~\ref{par3}.

It is interesting to emphasize that this line element corresponds to the
other of the vacuum solutions previously obtained in \cite{AyonBeato:2010tm}
for the same theory, but with a more restrictive choice of the coupling
constants. Notice that if we fix the coupling constant $\beta_3$ by demanding
the vanishing of the scalar strength we recover the black hole without a scalar
field of \cite{AyonBeato:2010tm}. In other words, this solution generalizes
the other vacuum example of a higher-dimensional Lifshitz black hole by
dressing it with a self-interacting nonminimally coupled scalar field.

All the previous solutions exist for generic values of the nonminimal
coupling parameter that include the minimal case $\xi=0$ in particular. In
the next section we provide an example which is necessarily nonminimal, but
this entails going beyond ansatz (\ref{gensolLif}).

%%%%%%%%%%%%%%%%%%%%%%%%%%%%%%%%%%%%%%%%%%%%%%%%%%%%%%%%%%%%%%%%%%%%%%%%%%%%%
\section{Last class of Lifshitz black holes \label{sec:lastc}}
%%%%%%%%%%%%%%%%%%%%%%%%%%%%%%%%%%%%%%%%%%%%%%%%%%%%%%%%%%%%%%%%%%%%%%%%%%%%%

The fifth class of Lifshitz black holes that does not fit within our ansatz
(\ref{gensolLif}) has a fixed value of the dynamical exponent $z=D$ and is
valid also for a precise value of the nonminimal coupling parameter
\[
\xi=\frac{(2D-1)(D-1)}{2(3D^2-2D+4)(D+1)}.
\]
The configuration in question reads
\begin{align}
ds^2={}&-\frac{r^{2D}}{l^{2D}}\!\left[1-M\!\left(\frac{l}{r}\right)^{\!\!2(D-1)}
-\alpha\sqrt{M}\!\left(\frac{l}{r}\right)^{\!\!D-1}\right]\!dt^2\nonumber\\
&+\frac{l^2}{r^2}
\!\left[1-M\!\left(\frac{l}{r}\right)^{\!\!2(D-1)}-\alpha\sqrt{M}
\!\left(\frac{l}{r}\right)^{\!\!D-1}\right]^{-1}\!\!dr^2\nonumber\\
&+\frac{r^2}{l^2}\sum_{i=1}^{D-2} dx_{i}^{2},\nonumber\\
\Phi(r)={}&\sqrt{\frac{2(D+1)(3D^2-2D+4)M}{\kappa(D-1)P_2(D)}}
\left(\frac{l}{r}\right)^{D-1},
\end{align}
where $\alpha$ is a coupling constant appearing in the potential and
$P_2(D)=6D^2-16D-1$. Indeed, this solution exists provided that the potential
and the coupling constants are given by
\begin{align*}
U(\Phi)={}&-\frac{(D-2)(D-1)^2P_2(D)}{4(D+1)(3D^2-2D+4)l^2}\Phi^{2}
\nonumber\\
&-\frac{\alpha
(2D-1)}{2l^2}\sqrt{\frac{(D-1)^{5}P_2(D)\kappa}
{2(D+1)(3D^2-2D+4)^3}}\,\Phi^{3}\\
&-\frac{(D-1)^3P_2(D)(D^3-4D^2+19D+2)\kappa}
{16(D+1)^2(3D^2-2D+4)^2l^2}\Phi^{4}, \\
\beta_{1}={}&\frac{(2D^3-6D^2+25D+3)l^2}{2(D-4)(D-1)^2P_2(D)},\\
\beta_{2}={}&-\frac{2(D+1)(D^2+3D-1)l^2}{(D-4)(D-1)^2P_2(D)},\\
\beta_{3}={}&\frac{3(D+1)l^2}{2(D-4)P_2(D)},\\
\lambda={}&-\frac {(D-1)(D-2)(18D^2-32D-1)}{4P_2(D)l^2}.
\end{align*}

The event horizon of this black hole is located at the radius
\begin{equation*}
r_h^{D-1}=l^{D-1}
\frac{\sqrt{M}}{2}\left(\alpha+\sqrt{\alpha^2+4}\right),
\end{equation*}
and in terms of this radius the quantities of interest to corroborate the
first law (\ref{firstlaw}), together with the validity of the Cardy-like
(\ref{generCardyd}) and Smarr (\ref{smarrf}) formulas, are given by
\begin{align*}
\mathcal{S}_\text{W}&=\frac{2\pi\Upsilon_3}{\kappa}
\left(\frac{r_h}{l}\right)^{D-2}\Omega_{D-2},\\
T&=\frac{(D-1)\sqrt{\alpha^2+4}}{2\pi l(\alpha+\sqrt{\alpha^2+4})}
\left(\frac{r_h}{l}\right)^{D},\\
\mathcal{M}_\text{bh}&=\frac{(D-2)\sqrt{\alpha^2+4}\Upsilon_3}
{2\kappa(\alpha+\sqrt{\alpha^2+4})}
\left(\frac{r_h}{l}\right)^{2(D-1)}\frac{\Omega_{D-2}}{l},\nonumber\\
\mathcal{M}_\text{sol}&=
\frac{-D\sqrt{\alpha^2+4}\Upsilon_3}{2\kappa(\alpha+\sqrt{\alpha^2+4})}
\!\!\left(\!\!\frac{\alpha+\sqrt{\alpha^2+4}}
{(D-1)\sqrt{\alpha^2+4}}\!\right)^{\!\!\!\frac{2(D-1)}{D}}\!\!
\frac{\Omega_{D-2}}{l},
\end{align*}
where the extensive coefficient is expressed as all the previous ones in
App.~\ref{App:Ecoeffs}.

%%%%%%%%%%%%%%%%%%%%%%%%%%%%%%%%%%%
\section{Conclusion}
%%%%%%%%%%%%%%%%%%%%%%%%%%%%%%%%%%%

Here, we have extended the work on dressed Lifshitz black holes done in three
dimensions in the case of a scalar field nonminimally coupled to New Massive
Gravity \cite{Ayon-Beato:2015jga}. Indeed, we have considered a gravity
theory given by the most general quadratic curvature corrections to Einstein
gravity supplemented by a source action describing a self-interacting
nonminimally coupled scalar field. For this theory, we have presented five
different classes of Lifshitz black hole solutions. Each solution is
specified with a particular self-interacting potential and certain
parametrization of the cosmological constant and the coupling constants
$\beta_n$ accompanying the different quadratic invariants.

Interestingly, some of the obtained solutions describe Lifshitz black hole
backgrounds that were known previously as part of the vacuum of the studied
theories \cite{AyonBeato:2010tm}, but for more restrictive elections of the
coupling constants. These restrictions are recovered from the presented
solutions in the limit of a vanishing scalar strength. Hence, the new
configurations constitute generalizations of these vacuum higher-dimensional
Lifshitz black holes that now turn out to be dressed by self-interacting
nonminimally coupled scalar fields.

We would like to stress that our work constitutes a new example that
underlines the importance played by the gravitational solitons in order to
describe the thermal properties of black holes with (un)usual asymptotics,
since they are indispensable to write the Cardy-like microscopic entropy
formula. In this spirit, it will be desirable to keep exploring this issue
from the holographic point of view. In particular, a promising work to be
done will consist of identifying or interpreting the role of the soliton in
the field theory side.

Another interesting aspect that has to do with these solutions concerns the
Smarr formula \cite{Smarr:1972kt}. Indeed, since all the solutions reported
here verify the higher-dimensional anisotropic Cardy-like formula
(\ref{generCardyd}) as well as the first law of thermodynamics
(\ref{firstlaw}), they will also satisfy an anisotropic higher-dimensional
version of the Smarr formula (\ref{smarrf}). This last formula is in perfect
accordance with the one discussed, for example, in Ref.~\cite{Brenna:2015pqa}
for different theories admitting Lifshitz black holes. In fact, the Smarr
formula is just a reflection of the proportionality between the entropy and
the black hole mass, a proportionality observed by all the extensive
thermodynamics quantities through the same coefficient $\Upsilon$,
generically defined in (\ref{eq:Upsilon}). For example, putting together the
Smarr formula (\ref{smarrf}) and the Cardy-like one (\ref{generCardyd}) it
follows the proportionality between the masses of the black holes and their
corresponding solitons
\begin{equation*}
\mathcal{M}_\text{bh}=-\frac{D-2}{z}(2\pi l\,T)^{\frac{z+D-2}{z}}
\mathcal{M}_\text{sol},
\end{equation*}
which is checked in all the new solutions.

Consequently, the condition that allows identifying Lifshitz critical points
of the theory where all the extensive thermodynamics quantities related to
Lifshitz black holes vanish is $\Upsilon=0$. For example, in the absence of a
scalar field the related condition is satisfied in the isotropic AdS case
($z=1$) for the four-dimensional critical gravity point
\cite{Lu:2011zk,Deser:2011xc}. Two of the new obtained configurations exist
just within the identified anisotropic points, providing examples of Lifshitz
critical black holes. It will be interesting to study if their perturbations
also have zero energy, with this additional criterion there will be no doubt
that these points define well-behaved theories as in the isotropic case.

It is evident that the emergence of the solutions presented here is
essentially due to the higher-order nature of the gravity theory together
with the nonminimal coupling of the scalar field to this gravity through the
conventional term $R\Phi^2$. One eventually can pursue the exploration on
this issue by studying other Lifshitz black hole solutions that may arise
from other nonminimal couplings such as those recently put in the spotlight
through the Horndeski Lagrangian \cite{Horndeski:1974wa}.

\begin{acknowledgments}
We thank J.~Oliva for his contribution at the initial stage of this work and
R.~Olea for valuable comments. This work has been partially funded by Grants
No.~A1-S-11548 from Conacyt and No.~1171475 from Fondecyt. M.B.\ is supported
by Grant Conicyt/Programa Fondecyt de Iniciaci\'on en Investigaci\'on
No.~11170037. F.C.\ is supported by the Alexander von Humboldt Foundation and
is grateful for the warm hospitality at Leibniz Universit\"at Hannover.
\end{acknowledgments}

%%%%%%%%%%%%%%%%%%%%%%%%%%%
\appendix
%%%%%%%%%%%%%%%%%%%%%%%%%%%

%%%%%%%%%%%%%%%%%%%%%%%%%%%%%%%%%%%%%%%%%%%%%%%%%%%%%%%%%%%%%%%%%%%%%%%%%%%%%%
\section{Expressions for the coefficients of the mass formulas
(\ref{massgensol}) and (\ref{massgensoliton}) \label{a1}}
%%%%%%%%%%%%%%%%%%%%%%%%%%%%%%%%%%%%%%%%%%%%%%%%%%%%%%%%%%%%%%%%%%%%%%%%%%%%%%

The dimensionless combinations of coupling constants appearing in the black
hole mass formula (\ref{massgensol}) are {\small
%\begin{widetext}
\begin{align*}
l^2\Psi_1 ={}& (2\chi+2z-D+2)[\chi^2-(3z+2D-4)\chi+2z^2\\
&+(D-2)(2z+D-1)]\beta_1+\{\chi^3-2(z+D-2)\chi^2\\
&-[z^2-(D-2)(3z+D-2)]\chi+(2z-D+2)\\
&\times(z^2+D-2)\}\beta_2+\{2\chi^3-(4z+3D-6)\chi^2\\
&-[2z^2-(D-2)(9z-2)]\chi+4z^3\\
&-2(D-2)(3z^2-3z+1)\}\beta_3,\frac{}{}\\
%\end{align*}
%\begin{align*}
l^2\Psi_2 ={}&
2\{2\chi^3-4(z+D-2)\chi^2-2[z^2-(D-1)(D-2)]\chi\\
&+(2z-D+2)[2z^2+(D-2)(2z+D-1)]\}
\beta_1\\
&+\{2\chi^3-(4z+3D-6)\chi^2-[2z^2-(D-2)\\
&\times(3z+D-2)]\chi+2(2z-D+2)(z^2+D-2)\}
\beta_2\\
&+4\{\chi^3-(2z+D-2)\chi^2-[z^2-(D-2)(3z-1)]\chi\\
&+2z^3-(D-2)(3z^2-3z+1)\}\beta_3+(D-2)l^2.
\end{align*}}%
%\end{widetext}

The corresponding dimensionless combinations appearing in turn in the soliton
mass formula (\ref{massgensoliton}) will be {\small
%\begin{widetext}
\begin{align*}
l^2\Xi_1 ={}& (4\chi-2z-D+6)[\chi^2-(3z+2D-4)\chi+2z^2\\
&+(D-2)(2z+D-1)]\beta_1+\{\chi^3-2(2z+D-2)\chi^2\\
&+[5z^2+(3D-10)z+D(D-2)]\chi\\
&-(2z+D-6)(z^2+D-2)\}\beta_2-\{(2z-D+6)\chi^2\\
&-[6z^2-(3D-2)z+6D-8]\chi\\
&+4z^3-2(D+2)z(z-1)+2D-12\}\beta_3,\frac{}{}\\
%\end{align*}
%\begin{align*}
l^2\Xi_2 ={}&
2\{2\chi^3-2(3z+2D-5)\chi^2+[6z^2+6(D-3)z\\
&+(D-2)(3D-7)]\chi-(2z+D-6)[2z^2\\
&+(D-2)(2z+D-1)]\}\beta_1+\{\chi^3-(3z+2D-4)\chi^2\\
&+[4z^2+(3D-10)z+D(D-2)]\chi-2(2z+D-6)\\
&\times(z^2+D-2)\}\beta_2-4[\chi^2-(z^2-z+D-1)\chi+2z^3\\
&-(D+2)z(z-1)+D-6]\beta_3-(\chi-2z-D+4)l^2.
\end{align*}}%
%\end{widetext}

%%%%%%%%%%%%%%%%%%%%%%%%%%%%%%%%%%%%%%%%%%%%%%%%%%%%%%%%%%%%%%%%%%%%%%%%%%%%%
\section{Parameters associated to the first class of solutions
(\ref{solutionfamily1}) \label{1st}}
%%%%%%%%%%%%%%%%%%%%%%%%%%%%%%%%%%%%%%%%%%%%%%%%%%%%%%%%%%%%%%%%%%%%%%%%%%%%%

The potential associated to the first class of the solutions of
Subsec.~\ref{1stclass} reads {\small
%\begin{widetext}
\begin{align*}
U(\Phi)\!=\!{}&\frac{16\bigl[2z^2+(D-2)(2z+D-1)\bigr]\xi-3(z+D-2)^2}{32l^2}
\Phi^2\\
&\!+\!\Bigl\{\!4\bigl[3z^2+(D+2)(D-2)\bigr]\xi-(z+D-2)^2\!\Bigr\}\!P_5(z;\xi)\\
&\!\times\!\kappa\Phi^4\bigg{/}\biggl(64\Bigl\{2(D-3)(D-4)(z+D-2)P_{4}(z)\beta_3\\
&\!-\!\bigl[3z^2+(D+2)(D-2)\bigr]P_{3}(z)l^2\Bigr\}\biggr),
\end{align*}}%
%\end{widetext}
while the coupling constants and the cosmological one are tied as follows
{\small
\begin{align*}
\beta_1 ={}&\bigl\{4[4\tilde{P}_6(z)\xi-(z+D-2)^2\tilde{P}_4(z)]\beta_3+(z+D-2)\\
&\times[4P_{3}(z)\xi-\tilde{\tilde{P}}_{3}(z)]l^2\bigr\}
\big{/}\bigl[2(z+D-2)P_5(z;\xi)\bigr],
%\label{beta1family1}
\frac{}{}\\
\beta_2 ={}& 4\bigl\{(z-1)(z+D-2)(3z^2+D^2-4)l^2-P_6(z;\xi)\beta_3\bigr\}
\frac{}{}\\
&\big{/}\bigl[(z+D-2)P_5(z;\xi)\bigr],\frac{}{}\\
\lambda ={}&
-\frac1{4l^2}\biggl(2z^2+(D-2)(2z+D-1)\\
&-\frac{4(D-3)(D-4)z(z+D-2)\beta_3}{l^2}\biggr),%\label{lambdafamily1}
\end{align*}}%
where for simplicity we have defined {\small
\begin{align*}
P_{3}(z)        \equiv{}& 9z^3-3(9D-14)z^2-(D-2)(5D-62)z\\
                        & -(D-2)(D^2-4D+36),\frac{}{}\\
\tilde{P}_{3}(z)\equiv{}& 6z^3-4(5D-8)z^2-3(D-2)(D-15)z\\
                        & -(D-2)(D^2-3D+26),\frac{}{}\\
P_{4}(z)        \equiv{}& 27z^4-36(3D-5)z^3-2(D-2)(5D-116)z^2\\
                        & -4(D-2)(D^2-D+30)z-(D+2)(D-2)^3,\frac{}{}\\
P_5(z;\xi)      \equiv{}& 4\left[2z^2+(D-2)(2z+D-1)\right]P_{3}(z)\xi\\
                        & -(z+D-2)^2\tilde{P}_{3}(z),\frac{}{}\\
\tilde{\tilde{P}}_{3}(z)
                \equiv{}& 15z^3-(19D-22)z^2-3(D-2)(D-18)z\\
                        & -(D-2)(D^2-4D+36),\frac{}{}\\
\tilde{P}_{4}(z)\equiv{}& 3z^4 - (7D-10)z^3 - (3D^2-2D-25)z^2\\
                        & -(D^3-14D^2+30D+4)z -(2D+7)(D-2)^2,\frac{}{}\\
\tilde{\tilde{P}}_{4}(z)
                \equiv{}& 9z^4 - 6(3D-4)z^3 - 8(D^2-10)z^2 + 2(D^3-4D^2\\
                        & +32D-80)z -(D-2)(D^3+2D^2-12D+24),\frac{}{}\\
\tilde{\tilde{\tilde{P}}}_{4}(z)
                \equiv{}& 6z^4 - 2(7D-10)z^3 - (13D^2-53D+34)z^2\\
                        & +2(D^3-D^2+D-16)z\\
                        & -(D-2)(D^3+D^2-10D+20),\frac{}{}\\
P_6(z;\xi)      \equiv{}& 4\left[2z^2+(D-2)(2z+D-1)\right]
                          \tilde{\tilde{P}}_4(z)\xi\\
                        & -(z+D-2)^{2}\tilde{\tilde{\tilde{P}}}_4(z),
                          \frac{}{}\\
\tilde{P}_6(z)  \equiv{}& 9z^6 - 3(3D-2)z^5 - (2D^2+90D-275)z^4\\
                        & -2(11D^3-46D^2-17D+172)z^3\\
                        & -(D-2)(7D^3-62D^2+212D-256)z^2\\
                        & -(D-2)(D-4)(D^3-8D^2+34D-76)z \\
                        & -(2D^2+7D-6)(D-2)^3.
\end{align*}}%

%%%%%%%%%%%%%%%%%%%%%%%%%%%%%%%%%%%%%%%%%%%%%%%%%%%%%%%%%%%%%%%%%%%%%%%%%%%%%
\section{Parameters associated to the second class of critical solutions
(\ref{3nd}) \label{par3}}
%%%%%%%%%%%%%%%%%%%%%%%%%%%%%%%%%%%%%%%%%%%%%%%%%%%%%%%%%%%%%%%%%%%%%%%%%%%%%

The self-interaction which ensures the existence of the critical solution
(\ref{3nd}) is given by {\small
\begin{align*}
U(\Phi) ={}& \frac{\tilde{P}_{2}(z;\xi)}{4l^2}\Bigl(2\Phi^2
+\bigl[(D-1)(2z-D-2)\xi-(z-1)\\
&\times(2z-D-1)\bigr]l^2\kappa\Phi^4
\big{/}\bigl[4(D-3)(D-4)z(z-D)\beta_3\\
&-(D-1)(2z-D-2)l^2\bigr]\Bigr),
\end{align*}}%
while the rest of the parameters are fixed as {\small
\begin{align*}
\beta_1 ={}&
\biggl(\Bigl\{4\bigl[3(D-2)z^4+(D^2-10)z^3+(D+2)(D-3)^2z^2\\
&-(D-2)(2D^2-8D+9)z-(D-2)(D-4)\bigr]\xi \\
&-4(z-1)^2\bigl[(D^2-D-3)z+D-4\bigr]\Bigr\}\beta_3+(z-1)\\
&\times\bigl[(D-2)(3z+D-4)\xi-(z-2)(D-1)\big]l^2
\biggr) \nonumber \\
&\big{/}\bigl[2(D-2)(z-1)(3z+D-4){\tilde{P}}_{2}(z;\xi)\bigr],  \\
\beta_2 ={}& -\biggl( 4\Bigl\{\bigl[2z^2+(D-2)(2z+D-1)\bigr]\\
&\times\bigl[6(D-2)z^2-(D^2-3D+8)z-2D+8\bigr]\xi \\
&-(z-1)^2\bigl[D(5D-11)z+2(D-1)(D-4)\bigr]\Bigr\}\beta_3\\
&-(D-1)(z-1)(2z-D-2)l^2\biggr) \\
&\big{/}\bigl[2(D-2)(z-1)(3z+D-4){\tilde{P}}_{2}(z;\xi)\bigr],\\
\lambda ={}& -\frac1{4l^2}\biggl(2z^2+(D-2)(2z+D-1)\\
&-\frac{4(D-3)(D-4)z(z+D-2)\beta_3}{l^2}\biggr).
\end{align*}}%

%%%%%%%%%%%%%%%%%%%%%%%%%%%%%%%%%%%%%%%%%%%%%%%%%%%%%%%%%%%%%%%%%%%%%%%%%%%%%
\section{Extensive coefficients\label{App:Ecoeffs}}
%%%%%%%%%%%%%%%%%%%%%%%%%%%%%%%%%%%%%%%%%%%%%%%%%%%%%%%%%%%%%%%%%%%%%%%%%%%%%

The extensive thermodynamic quantities of each presented solution are
proportional to a mutual coefficient encompassing the details to which the
theory is thermodynamically sensitive; i.e.\ the values of this coefficient
determine all the points in the parameter space of the theory probed by the
solution, which correspond to the same thermodynamic behavior. In this
appendix we present these extensive coefficients for all solutions. Since
they usually take long expressions, this not only allows us to write concise
expressions for the extensive thermodynamic quantities but also make evident
the aforementioned proportionality, that incidentally is the base of the
Smarr formula (\ref{smarrf}).

The dimensionless mutual coefficient appearing in the extensive thermodynamic
quantities of the first class of the solutions of Subsec.~\ref{1stclass} is
given by
\begin{align*}
l^2\Upsilon_1={}&\frac{(z+D-2)(3z-D-2)}{4P_{5}(z;\xi)}
\biggl(2(D-3)(D-4)\nonumber\\
&\times\!\Bigl\{\!
16(3z+D-2)\bigl[2z^2+(D-2)(2z+D-1)\bigr]\xi\nonumber\\
&-(5z+3D-6)(z+D-2)^2\Bigr\}\beta_3\nonumber\\
&-(z+D-2)\bigl[3z^2+(D-2)(D+2)\bigr]l^2\biggr),
\end{align*}
where the polynomial $P_{5}(z;\xi)$ was previously defined in App.~\ref{1st}.

The dimensionless extensive coefficient related to the second class of the
solutions of Subsec.~\ref{2stclass} is written as
\begin{align*}
l^2\Upsilon_2={}&\frac{8(D-2)(D-1)^2}{D+2}\beta_1
-\frac{4(D^3-D^2-2D-4)}{D+2}\beta_3\nonumber\\
&-\frac{(D-1)(D-2)(1-4\xi)l^2}{(D+2)\bigl[D-1-(5D-2)\xi\bigr]}.
\end{align*}

Finally, the dimensionless extensive coefficient of the last class in
Sec.~\ref{sec:lastc} is
\begin{equation*}
l^2\Upsilon_3=-\frac{2(D+1)(2D-1)\sqrt{\alpha^2+4}}
{P_2(D)(\alpha+\sqrt{\alpha^2+4})},
\end{equation*}
where the polynomial $P_2(D)$ is defined in the same section.

%%%%%%%%%%%%%%%%%%%%%%%%%%%

\end{document}